\RequirePackage{lineno}
\documentclass[prd,twocolumn,nofootinbib,superscriptaddress,floatfix]{revtex4}
\usepackage{cancel}
\usepackage{graphicx}
\usepackage{epsfig}
\usepackage{amsmath,amsfonts,amssymb}
\usepackage{t1enc}

\newcommand{\oh}{{\textstyle \frac{1}{2}}}

\newcommand{\ttbar}{t \bar t}
\newcommand{\afb}{A_{FB}}
\newcommand{\ac}{A_C}
\newcommand{\mttbar}{m_{t \bar t}}
\newcommand{\prp}{{\mbox{\tiny($'$)}}}

\begin{document}

\title{Collider-independent top quark forward-backward asymmetries: \\ standard model predictions}

\author{J. A. Aguilar-Saavedra}
\affiliation{Departamento de F\'{\i}sica Te\'orica y del Cosmos, Universidad de Granada,
18071 Granada, Spain}
\author{W. Bernreuther}
\affiliation{Institut f\"ur Theoretische Physik, RWTH Aachen University, 52056 Aachen, Germany}
\author{Z.-G. Si}
\affiliation{Department of Physics, Shandong University, Jinan, Shandong 250100, China}

\begin{abstract}
We compute, for top quark pair production at the Tevatron and the Large Hadron Collider, the collider-independent forward-backward asymmetries defined in~\cite{AguilarSaavedra:2012va} in the standard model at next-to-leading order in QCD, including also electromagnetic and weak corrections. 
\end{abstract}

\maketitle

\section{Introduction}

The charge asymmetry -- respectively forward-backward (FB) asymmetry -- in
$t \bar t$ production at the Fermilab Tevatron stands out as perhaps
the most  prominent anomaly that the data analysis of this collider
has left. The observable that has been mostly used is the $\ttbar $
rest frame asymmetry whose definition is  based on the difference $\Delta y = y_t - y_{\bar t}$ between the rapidities of the top quark and antiquark, which is invariant under boosts along the beam direction,
\begin{equation}
\afb = \frac{N(\Delta y > 0) - N(\Delta y < 0)}{N(\Delta y > 0) + N(\Delta y < 0)} \,,
\label{ec:afb}
\end{equation}
where $N$ denotes the respective number of $\ttbar$ events. The measurements of this asymmetry by the CDF~\cite{Aaltonen:2011kc,CDF8.7} and D0~\cite{Abazov:2011rq} Collaborations, which are in excess of the standard model (SM) expectations~\cite{Kuhn:1998kw,Antunano:2007da,Almeida:2008ug,Bernreuther:2010ny,Kidonakis:2011zn,Ahrens:2011uf,Hollik:2011ps,Kuhn:2011ri,Campbell:2012uf,Bernreuther:2012sx}, have triggered more than a hundred theory papers that explain this anomaly by new physics (see, for example, \cite{Djouadi:2009nb,Jung:2009jz,Cheung:2009ch,Nelson:2011us,Shu:2009xf,Barcelo:2011vk}). New physics explanations of the anomalous Tevatron asymmetry often predict new related effects at the Large Hadron Collider (LHC)
\cite{Cao:2011ew}, including the observation of new particles~\cite{Gresham:2011pa}. As yet, none of these effects have been found at the LHC. But, of course, this does not rule out the possibility that the Tevatron asymmetry results from new physics, telling us that, if anything, this new physics is perhaps not as simply modeled as by the hitherto existing proposals.

A closer test of the Tevatron excess is provided by the measurement of
the $\ttbar$ charge asymmetry at the LHC. While the Tevatron FB
asymmetry in Eq.~(\ref{ec:afb}) involves
  the rapidity difference $\Delta y$, the definition of the LHC charge
  asymmetry used by the CMS and ATLAS experiments employs the difference $\Delta |y| = |y_t| - |y_{\bar t}|$ between the absolute values of the top and anti-top rapidities in the laboratory (LAB) frame~\cite{Diener:2009ee},
\begin{equation}
\ac = \frac{N(\Delta |y| > 0) - N(\Delta |y| < 0)}{N(\Delta |y| > 0) + N(\Delta |y| < 0)} \,.
\label{ec:ac}
\end{equation}
This definition takes advantage of the fact that valence quarks $q=u,d$ have a larger average momentum fraction than antiquarks $\bar q$. This leads to a boost of the $t \bar t$ system along the direction of the incoming quark. Therefore, an excess of top quarks in this direction -- that is, a FB asymmetry in the center-of-mass (CM) frame of the initial partons -- leads to more $t$ than $\bar t$ quarks for large values $|y|$ of the (anti)top rapidity, while for small values of $|y|$ it is the other way around.
Current measurements of $\ac$  by the ATLAS~\cite{Aad:2012ug} and CMS~\cite{Chatrchyan:2011hk,Chatrchyan:2012xv} Collaborations have found agreement with the SM predictions. However, these results are per se not incompatible with the Tevatron measurements~\cite{AguilarSaavedra:2012va,Drobnak:2012cz,Alvarez:2012ca,Drobnak:2012rb}, since $\afb$ and $\ac$ are different observables that result from a different ``weighting'' of the ``intrinsic'' asymmetries $A_u$, $A_d$ in $u \bar u \to t \bar t$, $d \bar d \to t \bar t$, respectively. (Notice that $gg \to \ttbar$ does not contribute to $\afb$ and $\ac$.) In this way, models giving rise to different intrinsic asymmetries $A_u$, $A_d$, lead to different predictions for the relation between $\ac$ and $\afb$~\cite{AguilarSaavedra:2011hz}.

A direct test of the Tevatron anomaly has been proposed~\cite{AguilarSaavedra:2012va} that consists
in the extraction of the asymmetries $A_{u,d}$ from the measurement of the suitably binned asymmetries of Eqs.~(\ref{ec:afb}), (\ref{ec:ac}) at the Tevatron and LHC, respectively, and the subsequent comparison of the respective results. Their numerical values are nearly the same at both colliders, up to corrections that are much smaller than the experimental precision; thus their denomination as ``collider-independent''. The determination of the {\it same} quantities at the two colliders could shed light on the origin of the Tevatron anomalies and settle the apparent tension with the LHC measurements.

The asymmetries $A_{u,d}$ can be extracted from the FB and charge asymmetries in Eqs.~(\ref{ec:afb}), (\ref{ec:ac}) because they can be written, to a good approximation, as
\begin{eqnarray}
\afb & = & A_u F_u + A_d F_d \,, \notag \\
\ac & = & A_u F_u D_u + A_d F_d D_d \,,
\label{ec:AuAd}
\end{eqnarray}
provided we restrict ourselves to a narrow interval in the $\ttbar$ invariant mass $\mttbar$.
This will be shown below.\footnote{In fact, in   the derivation of Eqs.~(\ref{ec:AuAd}), it is the partonic squared CM energy $\hat s$ that has to be fixed, which differs in general
 from $m_{t\bar t}$. But fixing  $m_{t\bar t}$ instead of $\hat s$, which is required
 in applications to data analysis,
  is good enough for our purpose.}
The factors $F_q$ -- which differ at Tevatron and the LHC and also depend on the CM energy -- can be interpreted at leading order QCD as the fractions of $q \bar q$ initiated $\ttbar$ events.
   The  $D_q$ are ``dilution'' factors that take into account that in $\ttbar$ production at the LHC it often happens that the initial valence quark has smaller momentum fraction than the sea antiquark, thus leading to a dilution of the asymmetry generated at the partonic level. Both $F_q$ and 
 $D_q$ depend on the longitudinal velocity of the $t \bar t$ pair in the LAB frame,
\begin{equation}
\beta = \frac{|p_t^z + p_{\bar t}^z|}{E_t + E_{\bar t}} \,,
\end{equation}
where $E$, $p^z$ are the LAB-frame (anti)top energy and momentum along the beam direction, respectively.
On the other hand, for fixed $\hat s$, $A_{u,d}$ are $\beta$-independent.
In practice, where a  finite $m_{t\bar t}$ interval has to be used
instead  of $\hat s$, $A_{u,d}$ become mildly $\beta$-dependent.
This $\beta$-dependence can be weakened by imposing an upper cut on the
 transverse momentum $p_T^{t\bar t}$ of the $t\bar t$ pair. Hence, for a chosen
interval $m_L < m_{t\bar t} < m_U$, $A_{u,d}$ can be extracted from
a fit to the distributions $A_{FB}(\beta)$ and $A_C(\beta)$ measured
within this $m_{t\bar t}$ interval, as discussed in~\cite{AguilarSaavedra:2012va}, using the $F_q$ and $D_q$ factors computed in the SM, {\it e.g.} by Monte Carlo.

In this paper we calculate $A_{u,d}$ in the SM at next-to-leading order (NLO) in the QCD coupling, including also electromagnetic and weak corrections. To be precise, ``NLO'' refers in this paper to the computation of the numerators in Eqs.~(\ref{ec:afb}), (\ref{ec:ac}) to order $\alpha_s^3$ including the electroweak corrections of order $\alpha_s^2 \alpha$.
In the next section we derive Eqs.~(\ref{ec:AuAd}) in detail for the
SM at NLO. This derivation holds also if there are new physics
contributions to the asymmetries  (\ref{ec:afb}), (\ref{ec:ac}). 
In addition, we discuss the role of $gq$ contributions, which also
lead to an asymmetry at the LHC, 
 albeit very small in the SM. In section~\ref{sec:3} we present our numerical results.

\section{Derivations}
\label{sec:2}

The following derivations apply to the computation of the numerators of Eqs.~(\ref{ec:afb}), (\ref{ec:ac}) to NLO in the gauge couplings (see above). These numerators receive non-zero contributions only
 from terms in the squared  matrix elements that are asymmetric with
respect to the exchange of the $t$ and $\bar t$ momenta. As it is well-known, respective  contributions  $d\sigma^A$ only arise from the matrix elements of $q \bar q \to t
\bar t (g, \gamma)$, where $q=u,d$, and of $g q (\bar q) \to t \bar t q (\bar q)$. To NLO in the gauge couplings, the charge asymmetric terms
$d\sigma^A_{qg}$ are infrared-finite, while for $q \bar q$ initiated
$t\bar t$ production, the soft-gluon divergence that is present in
$d\sigma^A_{q \bar q}$ (virtual + soft) cancels against
  the corresponding divergence in
$\int d\sigma^A_{q \bar q}$ ($t {\bar t} g),$ and likewise for real
photon radiation.  To NLO in the gauge
couplings, the numerators of Eqs.~(\ref{ec:afb}), (\ref{ec:ac}) are free of  initial-state
collinear singularities -- i.e., no collinear counterterms are required
to this order. 

The NLO numerators are denoted by $N^1$ in the following.
For definiteness, we consider the denominators of Eqs.~(\ref{ec:afb}), (\ref{ec:ac}) to
leading order (LO) and denote them by $N^0$. Yet, alternatively, NLO
  denominators may be used, and the derivations are completely analogous to the ones presented here. Quantities without subindices imply a sum over all partonic sub-processes, whereas a subindex, if present, indicates the corresponding subprocess.
For brevity we label with superscripts $F,B$ the $\ttbar$ events with $\Delta y \gtrless 0$, respectively, and with superscripts $>,<$ the events with $\Delta |y| \gtrless 0$.

To NLO, the numerator of Eq.~(\ref{ec:afb}) receives
contributions from $q {\bar q} \to t {\bar t} (g,\gamma)$, $q=u,d$.
Top-quark pair production by $gg$ fusion is symmetric, and
  contributions to the numerator of Eq.~(\ref{ec:afb}) by $qg$, ${\bar q}g$
  processes are completely negligible at the Tevatron.
  Then, the FB asymmetry takes the form
\begin{eqnarray}
\afb & = & \frac{N_{u\bar u}^{1,F} - N_{u\bar u}^{1,B}}{N^0} + (u\to d) \notag \\
& = & \frac{N_{u\bar u}^{1,F} - N_{u\bar u}^{1,B}}{N_{u\bar u}^0} \frac{N_{u\bar u}^0}{N^0} + (u\to d) \notag \\
& \equiv & A_u F_u + A_d F_d \,.
\label{ec:AuAd-tev}
\end{eqnarray}
Likewise, the LHC charge asymmetry in Eq.~(\ref{ec:ac}) can be written as
\begin{eqnarray}
\ac & = & \frac{N_{u\bar u}^{1,>} - N_{u\bar u}^{1,<}}{N^0} +  \frac{N_{gu}^{1,>} - N_{gu}^{1,<}}{N^0} + (u\to d) \notag \\
& = & \frac{N_{u\bar u}^{1,>} - N_{u\bar u}^{1,<}}{N_{u\bar u}^0} \frac{N_{u\bar u}^0}{N^0} 
+ \frac{N_{gu}^{1,>} - N_{gu}^{1,<}}{N_{u\bar u}^0} \frac{N_{u\bar u}^0}{N^0} \notag \\
& &  + (u\to d) \,.
\label{ec:AuAd-lhc}
\end{eqnarray}
 The denominators $N^0$ in Eqs.~(\ref{ec:AuAd-tev}) and~(\ref{ec:AuAd-lhc}) are the (binned) LO
 QCD $t\bar t$ cross sections at the Tevatron and LHC,
 respectively. For ease of notation, we use the symbols $N^0$,
 $N_{q\bar q}^{1,F}$, etc., both for the Tevatron and the LHC.
 The SM contributions to the numerator of the charge asymmetry from $gq$ sub-processes are rather small~\cite{Kuhn:2011ri,Bernreuther:2012sx} and can be ignored.\footnote{If these contributions were measurable, it would be more adequate to use $N_{gq}^1$ for the relative normalization in Eq.~(\ref{ec:AuAd-lhc}), instead of $N_{q\bar q}^0$. On the other hand, our choice shows more clearly the relative size of the $gq$ asymmetries, compared to $q \bar q$.} (We will explicitly compute them in the next section.) Moreover, it will be shown below that, provided we restrict ourselves to a 
narrow $\hat s$ interval and to small values of 
 $p_T^{t\bar t}$, the differences $N_{q\bar q}^{1,>} - N_{q\bar q}^{1,<}$ are related to the differences $N_{q\bar q}^{1,F} - N_{q\bar q}^{1,B}$ (where ``forward'' and ``backward'' refer to the initial quark direction) by
\begin{equation}
N_{q\bar q}^{1,>} - N_{q\bar q}^{1,<} = D_q \left( N_{q\bar q}^{1,F} - N_{q\bar q}^{1,B} \right) \,.
\label{ec:dil1}
\end{equation}
Here $D_q$ is a so-called dilution factor, defined,  again for a narrow
  interval in $\hat s$,  by
\begin{equation} \label{dilfac}
D_q \equiv \frac{N(x_q > x_{\bar q}) - N(x_q < x_{\bar q})}{N(x_q > x_{\bar q}) + N(x_q < x_{\bar q})} \,.
\end{equation}
Using Eq.~(\ref{ec:dil1}) and neglecting
  the $gq$ contributions,  the charge asymmetry (\ref{ec:AuAd-lhc}) can be written as
\begin{eqnarray}
\ac & = &  \frac{N_{u\bar u}^{1,>} - N_{u\bar u}^{1,<}}{N_{u\bar u}^0} \frac{N_{u\bar u}^0}{N^0} + (u \to d)
\notag \\
& = & \frac{N_{u\bar u}^{1,F} - N_{u\bar u}^{1,B}}{N_{u\bar u}^0} \frac{N_{u\bar u}^0}{N^0} D_u + (u \to d)
\notag \\
& \equiv & A'_u F'_u D_u + A'_d F'_d D_d \,. \label{defaudlhc}
\end{eqnarray}
Here, we have put primes  on $A'_q$ and $F'_q$ to emphasize that these
quantities correspond to the LHC, while  the unprimed quantities refer
to the Tevatron. However, as it will be shown below, the asymmetries $A'_{u,d}$ are, for the
same narrow interval in $\hat s$, approximately equal to the Tevatron asymmetries
  $A_{u,d}$ defined in Eq.~(\ref{ec:AuAd-tev}).

We will first show the equality between $A_{u,d}$ and  $A_{u,d}'$. For the latter, the forward and backward directions are defined with respect to the initial quark momentum direction. (Of course this is impossible to tell event by event.) Then, we derive Eq.~(\ref{ec:dil1}). Our notation is as follows.
We denote by $x_i$, $i=1,2$ the momentum fractions of the initial partons, and $f_p (x_i)$ is the distribution function for parton $p$ in the proton with momentum fraction $x_i$.
The dependence of  the parton distribution functions (PDF) on the factorization scale $\mu_F$ is not exhibited. The (anti)proton 4-momenta at the Tevatron and LHC, respectively, are
   denoted by $P_{1,2}$, and $d\hat\sigma$ denotes the differential cross
section of a  partonic subprocess which includes the
corresponding phase-space measure  $d\Phi$  and flux factor. 

In the following, we consider binned asymmetries by restricting
 the partonic CM energy $\hat s$  
 to an interval $[\hat s_\text{min},\hat s_\text{max}]$. This
  is accomplished  by a factor
\begin{equation} \label{factorH}
\mathcal{H} = \theta(\hat s-\hat s_\text{min})\theta(\hat s_\text{max}-\hat s)
\end{equation}
in the integrals.

At the Tevatron, the FB asymmetries $A_{u,d}$ are defined with respect to the proton direction. Their numerators are, in terms of the proton PDF, 
\begin{widetext}
\begin{eqnarray}
N_{q\bar q}^{1,F} - N_{q\bar q}^{1,B} & = & \int dx_1 dx_2 \, f_q(x_1) f_{q}(x_2)
 d\hat \sigma^1 [ q(x_1 P_1) \bar q(x_2 P_2) ]  \left[\theta(\Delta y)-\theta(-\Delta y)\right] \mathcal{H} \notag \\
 & & + \int dx_1 dx_2 \, f_{\bar q}(x_1) f_{\bar q}(x_2)
d\hat \sigma^1 [ q(x_2 P_2) \bar q(x_1 P_1)] \left[\theta(\Delta y)-\theta(-\Delta y)\right] \mathcal{H}  \,,
\end{eqnarray}
\end{widetext}
where $d{\hat\sigma}^1$ denotes here the sum of the $q\bar q$
initiated NLO differential
cross sections for 2-particle
 and 3-particle final states, 
and $\Delta y$ is the difference of the 
  $t$ and $\bar t$ rapidities in the CM frame of the initial partons. 
 The second integral corresponds to events where the initial anti-quark comes from the proton and the quark from the anti-proton, and is much smaller than the first integral. (It amounts to a ``dilution'' of order $10^{-3}$ in the asymmetry.)
 Choosing $\hat s_\text{max}$ and $\hat s_\text{min}$ in  (\ref{factorH}) close enough to each other, 
the factor $\mathcal{H}$ fixes $\hat s$ within a suitably narrow interval, 
 in which $\int d\hat \sigma^1$, which are functions 
  of $\hat s$,  are nearly constant  and can then 
 be taken out of the $x_i$ integrals. Dropping the argument in $d\hat \sigma^1$ for brevity,
\begin{eqnarray}
N_{q\bar q}^{1,F} - N_{q\bar q}^{1,B} & = & \int d\hat \sigma^1 [\theta(\Delta y)-\theta(-\Delta y)] \notag \\
& & \times \int dx_1 dx_2 \, f_q(x_1) f_q(x_2) \mathcal{H} \,.
\end{eqnarray}
Notice that  $\Delta y$ is independent of $x_1$ and $x_2$.
  The same can be done for the LO denominators,
\begin{equation}
N_{q\bar q}^{0,F} + N_{q\bar q}^{0,B} = \int d\hat \sigma^0_{q \bar q}
   \int dx_i \, f_q(x_1) f_q(x_2) \mathcal{H} \,,
\end{equation}
where $d\sigma^0_{q \bar q}$ is 
 the LO differential cross section for $q \bar q \to \ttbar$,
so the asymmetries are
\begin{equation} \label{audtev}
A_q = \frac{ \int d\hat \sigma^1  \left[\theta(\Delta y)-\theta(-\Delta y) \right]}{\int d\hat \sigma^0 _{q \bar q} } \,.
\end{equation}
The numerators of the LHC
 ``FB'' asymmetries $A'_u$, $A'_d$ defined in  (\ref{defaudlhc}) are
\begin{widetext}
\begin{eqnarray}
N_{q\bar q}^{1,F} - N_{q\bar q}^{1,B} & = & \int dx_1 dx_2 \, f_q(x_1) f_{\bar q}(x_2)
 d\hat \sigma^1 [ q(x_1 P_1) \bar q(x_2 P_2) ] \left[\theta(\Delta y)-\theta(-\Delta y) \right] \mathcal{H} \notag \\
 & & + \int dx_1 dx_2 \, f_{\bar q}(x_1) f_q(x_2)
d\hat \sigma^1 [ q(x_2 P_2) \bar q(x_1 P_1)] \left[\theta(-\Delta y)-\theta(\Delta y) \right] \mathcal{H} \,, \label{nqqlhcfb}
\end{eqnarray}
\end{widetext}
where the ``forward'' and ``backward'' directions are defined with respect to the incoming quark direction (note the opposite signs in the arguments of the $\theta$ functions of the second integral). By rotational invariance, the
  second term is equal to the first one, so we can concentrate on the former. Taking again the phase-space integrated partonic cross sections
   out of the $x_i$ integrals, we have
\begin{eqnarray}
\oh \left( N_{q\bar q}^{1,F} - N_{q\bar q}^{1,B} \right) & = & \int d\hat \sigma^1 
\left[\theta(\Delta y)-\theta(-\Delta y) \right] \notag \\
& & \times \int dx_1 dx_2 \, f_q(x_1) f_{\bar q}(x_2) \mathcal{H} \,. \notag \\ 
\label{ec:NlhcFB}
\end{eqnarray}
The LO denominators  $N^0_{q\bar q}=  N^{0,F}_{q\bar q} + N^{0,B}_{q\bar q}$
  of   $A'_{u,d}$ are 
\begin{eqnarray}
\oh   \left(N^{0,F}_{q\bar q} + N^{0,B}_{q\bar q} \right)
     & = & \int d\hat \sigma^0_{q\bar q} \notag \\
& & \times \int dx_1 dx_2 \, f_q(x_1) f_{\bar q}(x_2) \mathcal{H} \, . \notag \\
\end{eqnarray}
Thus the LHC asymmetries $A'_{u,d}$ are, given for  fixed $\hat s$, by
\begin{equation} \label{audlhc}
A'_q = \frac{ \int d\hat \sigma^1  \left[\theta(\Delta y)-\theta(-\Delta y) \right]}{\int d\hat \sigma^0_{q\bar q}} \, .
\end{equation}
They are   equal to the ones at the Tevatron, Eq.~(\ref{audtev}).

Next we show under which conditions Eq.~(\ref{ec:dil1}) holds. 
 The contribution from $q \bar q$ initial states to the numerator of the 
  binned LHC charge asymmetry (\ref{ec:AuAd-lhc}) is
\begin{widetext}
\begin{eqnarray}
N_{q\bar q}^{1,>} - N_{q\bar q}^{1,<} & = & \int dx_1 dx_2\, f_q(x_1) f_{\bar q}(x_2)
d\hat \sigma^1 [ q(x_1 P_1) \bar q(x_2 P_2) ] \mathcal{H} \mathcal{E} \notag \\
 & & + \int dx_1 dx_2  \, f_{\bar q}(x_1) f_q(x_2)
d\hat \sigma^1 [ q(x_2 P_2) \bar q(x_1 P_1) ] \mathcal{H} \mathcal{E} \,,
\label{ec:Nlhc}
\end{eqnarray}
\end{widetext}
where the asymmetric terms are selected by the factor
\begin{equation}
\mathcal{E} = \theta(|y_t|-|y_{\bar t}|) - \theta(|y_{\bar t}|-|y_t|) \,.
\end{equation}
Here $y_t$, $y_{\bar t}$ are the top and anti-top rapidities in the laboratory (LAB) frame, respectively. Using rotational invariance,
  the first integral equals the second one, so we can concentrate on the former. We now perform a rotation-free boost
 to the $t \bar t$ rest frame. Using that the sign of the difference of 
  the $t$ and $\bar t$ rapidities is frame invariant, we
  obtain, with some algebra, that in the limit of $p_T^{t\bar t}\to 0$, 
\begin{equation} \label{Erelation}
\mathcal{E}\to [ \theta(x_1-x_2) - \theta(x_2-x_1) ]  [ \theta(\Delta 
 y) - \theta(-\Delta y) ] \, .
\end{equation}
Inserting   Eq.~(\ref{Erelation})   into Eq.~(\ref{ec:Nlhc}) we obtain
 that for $t\bar t$ events with sufficiently small $p_T^{t\bar t}$,
\begin{widetext}
\begin{eqnarray}
\oh (N_{q\bar q}^{1,>} - N_{q\bar q}^{1,<}) & = & \int dx_1 dx_2 \, f_q(x_1) f_{\bar q}(x_2)  \times \left\{
\theta(x_1-x_2) \theta(\Delta y) - \theta(x_1-x_2) \theta(-\Delta y) - \theta(x_2-x_1) \theta(\Delta y)
\right. \notag \\
& & \left.  + \theta(x_2-x_1) \theta(-\Delta y)
\right \}  \times d\hat \sigma^1 \mathcal{H} \,.
\end{eqnarray}
\end{widetext}
Again, the factor $\mathcal{H}$ fixes $\hat s$ within a suitably narrow interval, 
 in which the  $\int d\hat \sigma^1$ are nearly constant 
 and can be taken out of the $x_i$ integrals:
\begin{widetext}
\begin{eqnarray}
\oh (N_{q\bar q}^{1,>} - N_{q\bar q}^{1,<})  & = &
 \int d\hat \sigma^1 \theta(\Delta y) \int dx_1 dx_2 \, f_q(x_1) f_{\bar q}(x_2) \theta(x_1-x_2) \mathcal{H} \notag \\
& & - \int d\hat \sigma^1 \theta(-\Delta y) \int dx_1 dx_2 \, f_q(x_1) f_{\bar q}(x_2) \theta(x_1-x_2) \mathcal{H} \notag \\
& & -  \int d\hat \sigma^1 \theta(\Delta y) \int dx_1 dx_2 \, f_q(x_1) f_{\bar q}(x_2) \theta(x_2-x_1) \mathcal{H} \notag \\
& & + \int d\hat \sigma^1 \theta(-\Delta y) \int dx_1 dx_2 \, f_q(x_1) f_{\bar q}(x_2) \theta(x_2-x_1) \mathcal{H} \,. 
\label{ec:Nlhc2}
\end{eqnarray}
\end{widetext}
Now let us define
\begin{eqnarray}
d_q^r & \equiv & \frac{ \int dx_1 dx_2  \, f_q(x_1) f_{\bar q}(x_2) d\hat \sigma_{q\bar q} \theta(x_1-x_2) \mathcal{H}}{ \int dx_1 dx_2  \, f_q(x_1) f_{\bar q}(x_2) 
  d\hat \sigma_{q\bar q} \mathcal{H}} \notag \\
& = & \frac{ \int dx_1 dx_2 \, f_q(x_1) f_{\bar q}(x_2) \theta(x_1-x_2) \mathcal{H}}{ \int dx_1 dx_2 \, f_q(x_1) f_{\bar q}(x_2) \mathcal{H}} \,,
\notag \\
d_q^w & \equiv & \frac{ \int dx_1 dx_2 \, f_q(x_1) f_{\bar q}(x_2) d\hat \sigma_{q\bar q} \theta(x_2-x_1) \mathcal{H}}{ \int dx_1 dx_2  \, f_q(x_1) f_{\bar q}(x_2)
   d\hat \sigma_{q\bar q} \mathcal{H}} \notag \\
& = & \frac{ \int dx_1 dx_2 \, f_q(x_1) f_{\bar q}(x_2) \theta(x_2-x_1) \mathcal{H}}{ \int dx_1 dx_2 f_q(x_1) f_{\bar q}(x_2) \mathcal{H}} \,.
\label{dqrw}
\end{eqnarray}
Here $d\hat \sigma_{q\bar q}$ denotes  the (LO or NLO)  differential cross section
  for $q {\bar q} \to t \bar t X$. In the next section,
  we use  $d\hat \sigma^0_{q\bar q}$ in  (\ref{dqrw}).
  The cancellation of the  $d\hat \sigma_{q\bar q}$ in the ratios (\ref{dqrw})
 works, for fixed $\hat s$, also to NLO, because
   all the terms in
   $d\hat \sigma^1_{q\bar q}$, which is the sum of the contributions
  from the  tree-level term, virtual corrections, soft and hard
  gluon radiation, and the collinear counterterm, are convoluted with the same
 product of PDF.

Clearly, $d_q^r+d_q^w=1$. The integrals with respect to
 $x_i$ in Eq.~(\ref{ec:Nlhc2}) can be written in terms of $d_q^r$ and $d_q^w$, 
  resulting in
\begin{widetext}
\begin{eqnarray}
\oh (N_{q\bar q}^{1,>} - N_{q\bar q}^{1,<})  & = &
 \int d\hat \sigma^1 \theta(\Delta y) \; d_q^r \int dx_1 dx_2 f_q(x_1) f_{\bar q}(x_2) \mathcal{H} - \int d\hat \sigma^1 \theta(-\Delta y) \; d_q^r  \int dx_1 dx_2 f_q(x_1) f_{\bar q}(x_2) \mathcal{H} \notag \\
& & -  \int d\hat \sigma^1 \theta(\Delta y) \; d_q^w \int dx_1 dx_2 f_q(x_1) f_{\bar q}(x_2) \mathcal{H}  + \int d\hat \sigma^1 \theta(-\Delta y) \; d_q^w \int dx_1 dx_2 f_q(x_1) f_{\bar q}(x_2) \mathcal{H} \,.
\end{eqnarray}
\end{widetext}
Rearranging terms, we have 
\begin{eqnarray}
\oh (N_{q\bar q}^{1,>} - N_{q\bar q}^{1,<}) & = & (d_q^r - d_q^w)  \int dx_1 dx_2 \, f_q(x_1) f_{\bar q}(x_2) \notag \\
& & \times d\hat \sigma^1 [\theta(\Delta y)-\theta(-\Delta y)] \mathcal{H} \,.
\end{eqnarray}
Comparing with Eq.~(\ref{ec:NlhcFB}) we obtain Eq.~(\ref{ec:dil1}), i.e., 
\begin{equation} \label{finderiv}
N_{q\bar q}^{1,>} - N_{q\bar q}^{1,<} = (d_q^r - d_q^w) \left(N_{q\bar q}^{1,F} - N_{q\bar q}^{1,B} \right) \,,
\end{equation}
 where 
\begin{equation}
 d_q^r - d_q^w = D_q
\end{equation}
are  the dilution factors  introduced in Eqs.~(\ref{ec:dil1}),~(\ref{dilfac}).

 We recall that this derivation holds for fixed $\hat s$ and sufficiently small
  $p_T^{t\bar t}$. On the other hand, the formula  
    (\ref{finderiv}) and the resulting formula (\ref{defaudlhc})
 holds  for arbitrary values $0< \beta <1$ of the longitudinal velocity of
   the $t\bar t$ system.
 In practice, the requirement of fixed $\hat s$ must be replaced by
     choosing a reasonably narrow bin in the $t \bar t$ invariant mass $m_{t\bar t}$,
   i.e.,
\begin{equation}
\mathcal{H} \longrightarrow \theta(m_{t\bar t} -m_L)\theta(m_U - m_{t\bar t}) \, .
\end{equation}
This will be done in the numerical computations of the next section.
 Then the intrinsic asymmetries  $A^\prp_{u,d}$ will become
 $\beta$-dependent; that is, the formulae Eqs.~(\ref{audtev}) and (\ref{audlhc})
 do no longer apply -- for the computations one has to use instead the definitions of
$A^\prp_{u,d}$ given in Eqs.~(\ref{ec:AuAd-tev}) and~(\ref{defaudlhc}),
  respectively.
But we will show that, for a given $m_{t\bar t}$ bin,
 this $\beta$-dependence is rather mild in the SM to NLO. 
  More importantly, as the results below will
 signify, $A'_u$ and $A'_d$  remain equal  to $A_u$ and $A_d$,  
 respectively,  to a good approximation -- 
  even if no upper cut is imposed on  $p_T^{t\bar t}$.
 In addition we will show by numerical computation that neglecting
 the $gq$ contributions to the LHC 
 charge asymmetry $A_C$  in the formula (\ref{ec:AuAd-lhc}) is indeed justified, given
 the level of precision one aims at in applying Eq.(\ref{ec:AuAd-lhc}) to
  future data analysis.

\section{Numerical results}
\label{sec:3}

Our numerical calculations are based on the code described in~\cite{Bernreuther:2010ny,Bernreuther:2012sx}. We compute the binned asymmetries of Eqs.~(\ref{ec:afb}) and (\ref{ec:ac}), and the asymmetries $A_{u,d}$ (Tevatron) and $A'_{u,d}$ (LHC, 7 and 8 TeV)
for a sequence of intervals $m_L < m_{t\bar t} < m_U$. Within a specified interval 
  $[m_L, m_U]$, the
asymmetries $\afb$ and $\ac$ are computed, for
 $\beta$ bins of width $\Delta\beta$ = 0.2 for the Tevatron (i.e.,
 $0< \beta <0.2$, etc.) and $\Delta\beta$ = 0.1  for the LHC (i.e.,
 $0< \beta <0.1$, etc.).

In the numerators of Eqs.~(\ref{ec:afb}) and (\ref{ec:ac}) and of $A\prp_{u,d}$ we take into account the ${\cal O}(\alpha_s^3)$ QCD and the   ${\cal  O}(\alpha_s^2\alpha)$ electroweak corrections.
For definiteness, we evaluate the denominators of all asymmetries considered in this paper with  LO QCD
matrix elements, which is in the spirit of a consistent fixed-order perturbative expansion of
ratios  like Eqs.~(\ref{ec:afb}), (\ref{ec:ac}). The 
  fractions $F^\prp_q$ and the dilution factors $D_q$
are computed for the Tevatron and the LHC (7 and 8 TeV) using LO QCD matrix
 elements both in the numerators and denominators.
We  evaluate both the numerators and denominators of the binned asymmetries  
and of $F^\prp_q$ and $D_q$
with  NLO parton distribution functions.

As emphasized above,  the analysis for $A^\prp_{u,d}$
could also be done by replacing, on the left- and right-hand
sides of Eqs. (\ref{ec:AuAd-tev}) and (\ref{ec:AuAd-lhc}), the global normalization factors
$N^0$ (Tevatron and LHC) 
 by the respective NLO factors  $N^1$.

We use $m_t=173.1$ GeV (on-shell mass), the QED coupling
  $\alpha(m_Z)=0.008$, and the weak mixing angle
  $\sin^2\theta_W=0.23$.  We use the  CTEQ6.6M
 PDF \cite{Nadolsky:2008zw} and the respective value of
 $\alpha_s(m_Z)$ provided by this set. We put $\mu_R=\mu_F =\mu$, and
numerical results are given for $\mu= m_t/2, m_t$, and $2 m_t$.
 These scale choices are purely conventional.
In Ref.~\cite{AguilarSaavedra:2012va} the asymmetries $A^\prp_{u,d}$ were
  obtained for a benchmark new physics model using a two-parameter fit
  to the $\afb(\beta)$ and $\ac(\beta)$ distributions, mimicking the
  procedure that has to be 
  eventually performed with real data. That can be
  done, for the Tevatron and the LHC,  by minimizing 
\begin{eqnarray}
\chi^2 & = & \sum_i \frac{\left[\afb(\beta_i) - A_u \; F_u(\beta_i) - A_d \; F_d(\beta_i)\right]^2}{\left[d\afb(\beta_i)\right]^2} \,, \notag \\
\chi^2 & = & \sum_i \frac{\left[\ac(\beta_i) - A'_u \; F'_u D_u(\beta_i) - A'_d \; F'_d D_d(\beta_i) \right]^2}{\left[d\ac(\beta_i)\right]^2} \,, \notag \\
\label{twofit}
\end{eqnarray}
with respect to $A^\prp_u$ and $A^\prp_d$. Here $i$ labels
   the different $\beta$ bins and $d\afb(\beta_i)$, $d\ac(\beta_i)$ 
  are the statistical uncertainties of the binned asymmetries.
Unfortunately, this procedure requires extremely high Monte Carlo statistics in order to have the two-parameter fit converging to the {\it true} values.
 Especially at the LHC, the $\beta$-binned $A'_{u,d}$ are obtained from the ratio of a tiny asymmetry $\ac(\beta_i)$ over a small $F_q D_q (\beta_i)$ factor. Therefore, in order to save computing time, we calculate the asymmetries with a one-parameter fit, considering $u \bar u$ and $d \bar d$ contributions separately.  
The values of $A^\prp_{u,d}$ presented in Tables~\ref{tab:tev-1}~-~\ref{tab:lhc8-1} below are obtained from
  a one-dimensional least squares parameter fit,
\begin{eqnarray}
A_q & = & \sum_i \frac{\afb^{q\bar q}(\beta_i)}{\left[d\afb^{q\bar q}(\beta_i)\right]^2} \left/  \sum_i  \frac{1}{\left[d\afb^{q\bar q}(\beta_i)\right]^2} \right. \,, \notag \\
A'_q & = & \sum_i \frac{\ac^{q\bar q}(\beta_i) / D_q(\beta_i)}{\left[d\ac^{q\bar q}(\beta_i)\right]^2} \left/   \sum_i  \frac{1}{\left[d\ac^{q\bar q}(\beta_i)\right]^2} \right. \,,
\label{ec:afit}
\end{eqnarray} 
where the superscripts of $\afb$, $\ac$ indicate 
 that we restrict the calculation to these specific sub-processes,
 eventually including $gq$ contributions as well. We will demonstrate
 below  the consistency of both 
 methods by showing that the values of $A^\prp_{u,d}$ calculated using 
either  (\ref{twofit}) or (\ref{ec:afit}) agree very well within the 
 expected experimental uncertainties. The one-parameter fit is more precise.

We estimate the statistical uncertainty of the $\beta$-binned asymmetries by taking an integrated luminosity of 20 fb$^{-1}$ for the Tevatron and 10 (30) fb$^{-1}$ for
  the LHC with 7 (8) TeV. This corresponds to an eventual combination 
 of results from both experiments at the Tevatron and LHC,
 respectively.  A selection efficiency 
   of $\sim  25\%$ is taken for the semileptonic $t \bar t$ decay channels,
similar to that found in the experimental 
 analyses~\cite{Aad:2012ug,Chatrchyan:2011hk}.
 This results in a combined efficiency factor of about 7\%. The results for the Tevatron are collected in Table~\ref{tab:tev-1}  and for the LHC in Tables~\ref{tab:lhc7-1} and \ref{tab:lhc8-1}, without and with an upper cut $p_T^{\ttbar} < 30$ GeV.

\begin{table}[htb]
\begin{center}
\caption{Asymmetries $A_{u,d}$ at the Tevatron, without  $p_T^{\ttbar}$ cut (top) and for
$p_T^{\ttbar} < 30$ GeV (bottom).}
\label{tab:tev-1}
\vspace{2mm}
\begin{tabular}{lcccccc}
\hline \hline
 & \multicolumn{2}{c}{$\mu= m_t$} & \multicolumn{2}{c}{$\mu= 2 m_t$} & \multicolumn{2}{c}{$\mu= m_t/2 $}\\
$\mttbar$ [GeV] & $A_u$ & $A_d$ & $A_u$ & $A_d$ & $A_u$ & $A_d$ \\
\hline
$ < 400$      & 0.058 & 0.039 & 0.054 & 0.036 & 0.061 & 0.044
\\
$400 - 450$ & 0.096 & 0.066 & 0.091 & 0.060 & 0.102 & 0.073
\\
$450 - 500$ & 0.123 & 0.086 & 0.116 & 0.079 & 0.131 & 0.095
\\
$500 - 550$ & 0.145 & 0.102 & 0.137 & 0.092 & 0.154 & 0.113
\\
$550 - 600$ & 0.164 & 0.115 & 0.156 & 0.106 & 0.176 & 0.128
\\
\hline \hline
\end{tabular}

\vspace{5mm}

\begin{tabular}{lcccccc}
\hline \hline
& \multicolumn{2}{c}{$\mu= m_t$} & \multicolumn{2}{c}{$\mu= 2 m_t$} & \multicolumn{2}{c}{$\mu= m_t/2 $}\\
$\mttbar$  [GeV] & $A_u$ & $A_d$ & $A_u$ & $A_d$ & $A_u$ & $A_d$ \\
\hline
$< 400$       & 0.069 & 0.046 & 0.065 & 0.042 & 0.075 & 0.051
\\
$400 - 450$ & 0.117 & 0.078 & 0.110 & 0.071 & 0.126 & 0.087
\\
$450 - 500$ & 0.150 & 0.101 & 0.141 & 0.092 & 0.161 & 0.113
\\
$500 - 550$ & 0.178 & 0.120 & 0.167 & 0.109 & 0.191 & 0.135
\\
$550 - 600$ & 0.201 & 0.137 & 0.190 & 0.125 & 0.217 & 0.153
\\
\hline \hline
\end{tabular}
\end{center}
\end{table}


\begin{table}[htb]
\begin{center}
\caption{Asymmetries $A'_{u,d}$ at the LHC (7 TeV), without  $p_T^{\ttbar}$ cut (top) and for
$p_T^{\ttbar} < 30$ GeV (bottom).}
\label{tab:lhc7-1}
\vspace{2mm}
\begin{tabular}{lcccccc}
\hline \hline
& \multicolumn{2}{c}{$\mu= m_t$} & \multicolumn{2}{c}{$\mu= 2 m_t$} & \multicolumn{2}{c}{$\mu= m_t/2 $}\\
$\mttbar$ [GeV]  & $A'_u$ & $A'_d$ & $A'_u$ & $A'_d$ & $A'_u$ & $A'_d$ \\
\hline
$ < 400$      & 0.055 & 0.038 & 0.052 & 0.035 & 0.059 & 0.042
\\
$400 - 450$ & 0.089 & 0.060 & 0.084 & 0.055 & 0.096 & 0.066
\\
$450 - 500$ & 0.112 & 0.077 & 0.106 & 0.070 & 0.120 & 0.085
\\
$500 - 550$ & 0.128 & 0.083 & 0.120 & 0.076 & 0.136 & 0.092
\\
$550 - 600$ & 0.142 & 0.093 & 0.134 & 0.085 & 0.151 & 0.101
\\
$600 - 650$ & 0.155 & 0.103 & 0.146 & 0.093 & 0.165 & 0.113
\\
$650 - 700$ & 0.164 & 0.110 & 0.156 & 0.102 & 0.177 & 0.122
\\
$700 - 750$ & 0.176 & 0.119 & 0.165 & 0.104 & 0.185 & 0.129
\\
$750 - 800$ & 0.182 & 0.118 & 0.170 & 0.107 & 0.195 & 0.131
\\
\hline \hline
\end{tabular}

\vspace{5mm}

\begin{tabular}{lcccccc}
\hline \hline
& \multicolumn{2}{c}{$\mu= m_t$} & \multicolumn{2}{c}{$\mu= 2 m_t$} & \multicolumn{2}{c}{$\mu= m_t/2 $}\\
$\mttbar$ [GeV] & $A'_u$ & $A'_d$ & $A'_u$ & $A'_d$ & $A'_u$ & $A'_d$ \\
\hline
$< 400$       & 0.071 & 0.054 & 0.068 & 0.047 & 0.077 & 0.059
\\
$400 - 450$ & 0.115 & 0.078 & 0.108 & 0.071 & 0.124 & 0.087
\\
$450 - 500$ & 0.149 & 0.103 & 0.140 & 0.093 & 0.160 & 0.114
\\
$500 - 550$ & 0.170 & 0.110 & 0.159 & 0.100 & 0.183 & 0.124
\\
$550 - 600$ & 0.193 & 0.128 & 0.180 & 0.116 & 0.209 & 0.144
\\
$600 - 650$ & 0.211 & 0.143 & 0.197 & 0.127 & 0.227 & 0.159
\\
$650 - 700$ & 0.229 & 0.153 & 0.215 & 0.140 & 0.247 & 0.169
\\
$700 - 750$ & 0.245 & 0.165 & 0.228 & 0.148 & 0.261 & 0.184
\\
$750 - 800$ & 0.262 & 0.177 & 0.244 & 0.158 & 0.282 & 0.194
\\
\hline \hline
\end{tabular}

\end{center}
\end{table}


\begin{table}[htb]
\begin{center}
\caption{Asymmetries $A'_{u,d}$ at the LHC (8 TeV), without  $p_T^{\ttbar}$ cut (top) and for
$p_T^{\ttbar} < 30$ GeV (bottom).}
\label{tab:lhc8-1}
\vspace{2mm}
\begin{tabular}{lcccccc}
\hline \hline
$\mttbar$ [GeV] & \multicolumn{2}{c}{$\mu= m_t$} & \multicolumn{2}{c}{$\mu= 2 m_t$} & \multicolumn{2}{c}{$\mu= m_t/2 $}\\
& $A'_u$ & $A'_d$ & $A'_u$ & $A'_d$ & $A'_u$ & $A'_d$ \\
\hline
$< 400$       & 0.055 & 0.038 & 0.052 & 0.035 & 0.059 & 0.042
\\
$400 - 450$ & 0.088 & 0.058 & 0.083 & 0.053 & 0.094 & 0.065
\\
$450 - 500$ & 0.111 & 0.075 & 0.105 & 0.070 & 0.118 & 0.084
\\
$500 - 550$ & 0.126 & 0.081 & 0.119 & 0.075 & 0.135 & 0.091
\\
$550 - 600$ & 0.139 & 0.089 & 0.131 & 0.081 & 0.147 & 0.097
\\
$600 - 650$ & 0.153 & 0.102 & 0.144 & 0.094 & 0.163 & 0.113
\\
$650 - 700$ & 0.161 & 0.107 & 0.153 & 0.099 & 0.172 & 0.120
\\
$700 - 750$ & 0.172 & 0.115 & 0.162 & 0.106 & 0.185 & 0.129
\\
$750 - 800$ & 0.177 & 0.115 & 0.167 & 0.106 & 0.190 & 0.131
\\
\hline \hline
\end{tabular}

\vspace{5mm}

\begin{tabular}{lcccccc}
\hline \hline
& \multicolumn{2}{c}{$\mu= m_t$} & \multicolumn{2}{c}{$\mu= 2 m_t$} & \multicolumn{2}{c}{$\mu= m_t/2 $}\\
$\mttbar$ [GeV] & $A'_u$ & $A'_d$ & $A'_u$ & $A'_d$ & $A'_u$ & $A'_d$ \\
\hline
$< 400$       & 0.073 & 0.040 & 0.068 & 0.050 & 0.078 & 0.061
\\
$400 - 450$ & 0.115 & 0.074 & 0.108 & 0.070 & 0.123 & 0.086
\\
$450 - 500$ & 0.147 & 0.101 & 0.139 & 0.092 & 0.159 & 0.113
\\
$500 - 550$ & 0.170 & 0.118 & 0.159 & 0.102 & 0.183 & 0.124
\\
$550 - 600$ & 0.191 & 0.128 & 0.179 & 0.113 & 0.205 & 0.141
\\
$600 - 650$ & 0.209 & 0.148 & 0.197 & 0.126 & 0.227 & 0.157
\\
$650 - 700$ & 0.229 & 0.157 & 0.215 & 0.139 & 0.246 & 0.168
\\
$700 - 750$ & 0.242 & 0.164 & 0.225 & 0.147 & 0.261 & 0.186
\\
$750 - 800$ & 0.258 & 0.174 & 0.243 & 0.155 & 0.277 & 0.193
\\
\hline \hline
\end{tabular}

\end{center}
\end{table}

\begin{figure*}[htb]
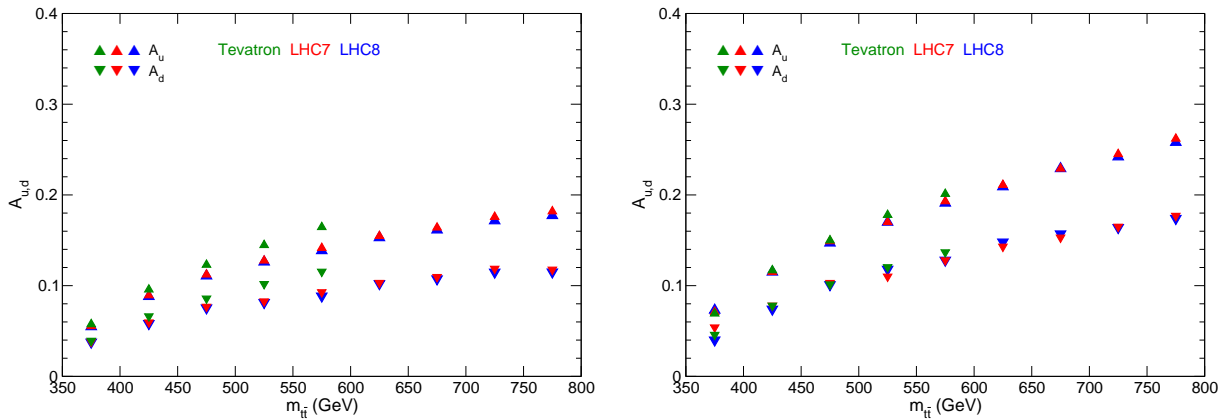

\begin{tabular}{ccc}
\epsfig{file=fig1a.eps,height=5.5cm,clip=} & \quad
\epsfig{file=fig1b.eps,height=5.5cm,clip=}
\end{tabular}
\caption{Summary of asymmetries for the Tevatron and the LHC (7 and 8 TeV), for $\mu=m_t$, without $p_T^{\ttbar}$ cut (left) and for $p_T^{\ttbar} < 30$ GeV (right). The data correspond to the first two columns of Tables~\ref{tab:tev-1}--\ref{tab:lhc8-1}.}
\label{fig:a-mtt}
\end{figure*}

These tables show that the SM values of $A_{u,d}$ and $A'_{u,d}$ computed for the Tevatron and the LHC, respectively, are in quite good agreement, which is remarkable given the difference of roughly one order of magnitude between the predictions for the inclusive asymmetries $\afb$ and $\ac$. For illustration, in Fig.~\ref{fig:a-mtt} the intrinsic asymmetries in Tables~\ref{tab:tev-1}--\ref{tab:lhc8-1}  are displayed, for $\mu =m_t$, as functions of $\mttbar$. For the Tevatron, only bins with $\mttbar < 600$ GeV are included.
These plots show the increase
of these asymmetries with  increasing $\mttbar$ and, furthermore, that the upper
cut on  $p_T^{\ttbar}$ reduces the slight difference between the Tevatron and LHC asymmetries, making them nearly equal. We remark that the differences exhibited in the left plot are irrelevant from an experimental point of view, as we shall see below. We also point out that
the differences between the Tevatron and LHC results originate from Monte Carlo statistics to some extent, as it can be noticed from the fact that the smooth increase of the asymmetries with $\mttbar$ is modulated by small fluctuations.

As we have emphasized before, at the LHC  the contributions from $gq$ processes to
$A'_{u,d}$ are quite small. This can be shown, for example, by  using in the second of Eqs.~(\ref{ec:afit})
   the SM values of $\ac^{q\bar q+gq}(\beta_i)$
  instead  of  $\ac^{q\bar q}(\beta_i)$. The resulting  intrinsic asymmetries are presented in Table \ref{tab:lhc7-2}
 for 7 TeV, without $p_T^{\ttbar}$ cut. Comparison with the
 corresponding numbers
  of Table~\ref{tab:lhc7-1} shows that the differences are
  negligible. The differences are  even further reduced if an upper cut on  $p_T^{\ttbar}$ is applied.

\begin{table}[htb]
\begin{center}
\caption{Asymmetries $A'_{u,d}$ at the LHC (7 TeV), without  $p_T^{\ttbar}$ cut, including $gq$ contributions.}
\label{tab:lhc7-2}
\vspace{2mm}
\begin{tabular}{lcccccc}
\hline \hline
& \multicolumn{2}{c}{$\mu= m_t$} & \multicolumn{2}{c}{$\mu= 2 m_t$} & \multicolumn{2}{c}{$\mu= m_t/2 $}\\
$\mttbar$ [GeV]  & $A'_u$ & $A'_d$ & $A'_u$ & $A'_d$ & $A'_u$ & $A'_d$ \\
\hline
$ < 400$      & 0.056 & 0.038 & 0.053 & 0.035 & 0.060 & 0.042
\\
$400 - 450$ & 0.091 & 0.060 & 0.086 & 0.055 & 0.097 & 0.067
\\
$450 - 500$ & 0.115 & 0.078 & 0.108 & 0.071 & 0.123 & 0.087
\\
$500 - 550$ & 0.132 & 0.085 & 0.124 & 0.078 & 0.141 & 0.095
\\
$550 - 600$ & 0.147 & 0.097 & 0.139 & 0.088 & 0.157 & 0.106
\\
$600 - 650$ & 0.161 & 0.108 & 0.152 & 0.097 & 0.172 & 0.118
\\
$650 - 700$ & 0.172 & 0.115 & 0.163 & 0.107 & 0.186 & 0.129
\\
$700 - 750$ & 0.184 & 0.125 & 0.172 & 0.110 & 0.195 & 0.136
\\
$750 - 800$ & 0.191 & 0.125 & 0.178 & 0.113 & 0.205 & 0.139
\\
\hline \hline
\end{tabular}
\end{center}
\end{table}

\begin{figure*}[htb]
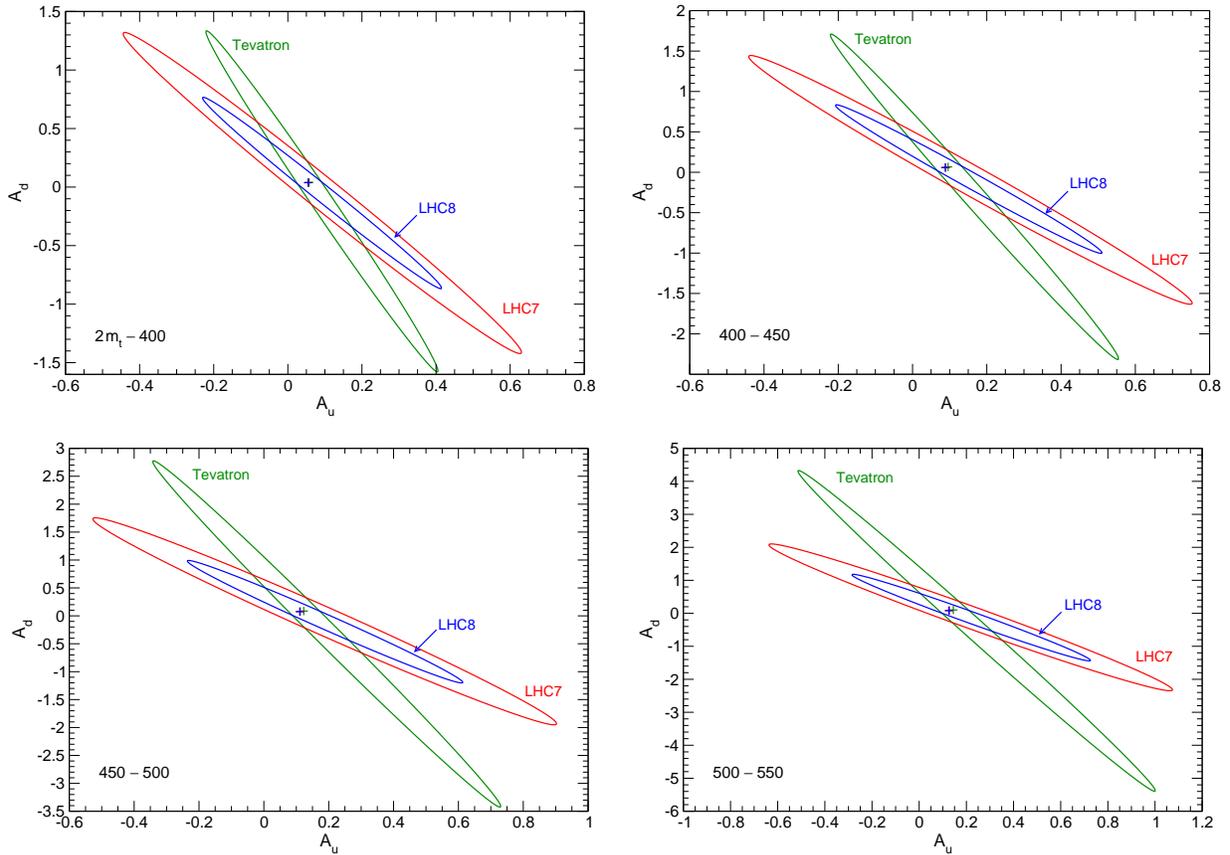

\begin{tabular}{ccc}
\epsfig{file=fig2a.eps,height=5.5cm,clip=} & \quad
\epsfig{file=fig2b.eps,height=5.5cm,clip=} \\[2mm]
\epsfig{file=fig2c.eps,height=5.5cm,clip=} & \quad
\epsfig{file=fig2d.eps,height=5.5cm,clip=}
\end{tabular}
\caption{Comparison between asymmetries obtained with two- and one-dimensional fits, for $\mu = m_t$, without $p_T^{\ttbar}$ cut. The mass range in the lower left corner indicates the $\mttbar$ bin (in GeV).
The ellipses represent the 68\% CL regions for the two-dimensional fit,
where only statistical uncertainties are considered.}
\label{fig:a2d}
\end{figure*}

Next, we check the equivalence of the two- and one-dimensional fits in
determining the intrinsic asymmetries. In Fig.~\ref{fig:a2d} 
  these asymmetries are plotted for the first four $\mttbar$ bins. The dots
represent the values calculated with one-parameter fits (these are the
numbers given in Tables~\ref{tab:tev-1}-\ref{tab:lhc8-1}), and the ellipses are the two-dimensional 68\% confidence level (CL) regions from the 2-dimensional fit, where the centre is the best-fit value giving the minimum $\chi^2_\text{min}$ and the border corresponds to $\chi^2 = \chi^2_\text{min} + 2.3$, including statistical uncertainties only and assuming a perfect reconstruction of the $t$ and $\bar t$ momenta. From these plots, it is also clear that the slight differences between the Tevatron and LHC asymmetries are irrelevant as to the anticipated experimental uncertainty. This justifies the ansatz of extracting the same quantities from two different sets of data.

So far we have determined the intrinsic asymmetries $A^\prp_{u,d}$ 
by simulating the proposed fitting procedure with SM data: we have
computed, for various $\mttbar$ bins, the binned asymmetries 
$\afb(\beta_i)$, $\ac(\beta_i)$, the fractions $F^\prp_q$ and the dilution
factors $D_q$ in the SM and performed the fits using Eqs.~(\ref{ec:afit}) and
(\ref{twofit}). This leads, by definition, to constant
 $A^\prp_{u,d}$ for each $\mttbar$ bin. It remains to show that 
 this is an acceptable procedure -- i.e., that the $\beta$-binned intrinsic asymmetries are
 only mildly $\beta$-dependent within the $\mttbar$ bins chosen above, as it was claimed.
This is shown in Fig.~\ref{fig:Audbeta} for the first $\mttbar$ bin. (For the other $\mttbar$ bins the behavior is quite similar.) This variation can be compared, for example, with an increase in $F'_u$ by a factor of 3 and $D_u$ ($D_d$) by factors  of 20 (40), between the bins $0<\beta<0.1$ and $0.9<\beta<1$.
These results corroborate the assumption of constant $A^\prp_{u,d}$, especially when an upper cut  on $p_T^{\ttbar}$ is used. The LHC results shown in the plot in the right panel of Fig.{\ref{fig:Audbeta}
exhibit some statistical fluctuations, which have some effect on the resulting
 fit values of of $A'_{u,d}$.

\begin{figure*}[htb]
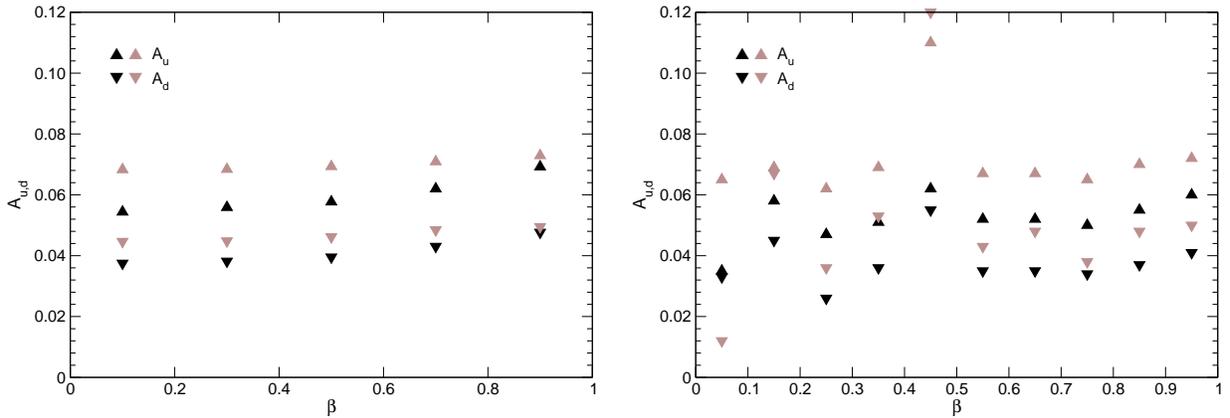

\begin{tabular}{ccc}
\epsfig{file=fig3a.eps,height=5.5cm,clip=} & \quad
\epsfig{file=fig3b.eps,height=5.5cm,clip=}
\end{tabular}
\caption{Dependence of $A_{u,d}$ on $\beta$ in the SM, for $\mttbar <
  400$ GeV at the Tevatron (left) and the LHC with 7 TeV (right). Black and  gray triangles correspond to no
  $p_T^{\ttbar}$ cut and cut $p_T^{\ttbar} < 30$ GeV, respectively.}
\label{fig:Audbeta}
\end{figure*}

Finally we comment on the importance of the SM electroweak
  contributions \cite{Hollik:2011ps,Kuhn:2011ri,Bernreuther:2012sx} of
order $\alpha_s^2\alpha$ to the charge asymmetries
   $A_{u,d}$ and $A'_{u,d}$. The dominant contributions are due to the
   photonic corrections \cite{Hollik:2011ps} whose size with respect to
the pure QCD asymmetries is roughly given by the ratio
$R_q=(36 Q_q Q_t/5)(\alpha/\alpha_s)$. That is, these QED  contributions
amount to a positive correction of $\sim 25\%$ for $A^\prp_u$, while
they  are negative, $\sim -13\%$ for $A^\prp_d$.
The pure QCD contributions to $A^\prp_{u,d}$ may of  course be
computed also with one of the generally
 available NLO QCD Monte Carlo programs,
e.g. with the codes of \cite{Campbell:2012uf,Frixione:2008ym}.
One should keep in mind, however, that in  Monte Carlo computations one
normalizes the asymmetries with NLO QCD denominators,  while we have
used denominators computed at LO QCD.

\section{Conclusions}

The formulae (\ref{ec:AuAd}) allow to extract the intrinsic forward-backward asymmetries  $A_{u,d}$ and $A'_{u,d}$ from the Tevatron $\ttbar$ forward-backward asymmetry $\afb$ and the LHC charge asymmetry $\ac$, respectively, if measured in suitably chosen bins of  the $t\bar t$ invariant mass $\mttbar$ and longitudinal velocity $\beta$ of the $t\bar t$ system. We have shown under which conditions Eqs.~(\ref{ec:AuAd}) hold in the SM to NLO in the gauge couplings. Our derivations apply of course also to possible new physics contributions to these asymmetries. In particular, we have shown within the SM that the intrinsic asymmetries are indeed  collider-independent and, furthermore,  only mildly $\beta$-dependent for suitably narrow $\mttbar$ bins, especially if an upper cut on the $p_T^{\ttbar}$ of the $t\bar t$ samples is applied. This corroborates the proposal of~\cite{AguilarSaavedra:2012va}  to use (\ref{ec:AuAd}) with constant $A_u$ and $A_d$ for performing a two-parameter fit to the respective Tevatron and LHC data. 

In order to  apply Eqs.~(\ref{ec:AuAd}) to data analysis, one has to compute the fractions $F_{q}$ and factors  $D_q$ in the SM,\footnote{The present knowledge about the (differential) hadronic $\ttbar$ production cross section implies that  possible new physics contributions to these functions can be neglected.}
 either to LO or NLO QCD  for a specific PDF set  and for chosen values of the renormalization and factorization scales. The outcome of the fits depend, of course,  on these choices. Needless to say, this
would be extremely valuable information. It has previously been shown, on a model-independent basis~\cite{AguilarSaavedra:2012va} as well as for specific new physics models~\cite{Drobnak:2012cz,Alvarez:2012ca,Drobnak:2012rb}, that current measurements of $\afb$ at the Tevatron and $\ac$ at the LHC are compatible. Therefore, the comparison of the measured values $A_{u,d}^\text{\tiny exp}$ with their SM predictions $A_{u,d}^\text{\tiny SM}$ would reveal whether or not there is agreement with the SM -- and if  a deviation would be found, it would show whether it is located in $A_u$ or $A_d$, or in both.
This could be achieved by a combination
 of the results from measurements at the Tevatron and the LHC (and by
eventually testing whether they are consistent), and would be a big step
forward in pinning down the origin of the new physics contribution(s),
if there are any.
The experimental determination of $A_u$ and $A_d$ will certainly be a challenge, as it will involve in general a 3-dimensional unfolding of the data with respect to $\mttbar$, $\beta$, and the (anti)top rapidity, but it is certainly worth the effort, for the aim of resolving this puzzle.

\subsubsection*{Acknowledgements}
The work of J.A.A.S. has been supported by MICINN by projects FPA2006-05294 and FPA2010-17915, Junta de Andaluc\'{\i}a (FQM 101, FQM 03048 and FQM 6552) and Funda\c c\~ao
para a Ci\^encia e Tecnologia~(FCT) project CERN/FP/123619/2011.
The work of   W.B.  was supported by DFG, SFB TR9 and that of Z.G. Si  by NSFC and by Natural Science Foundation of Shandong Province.


\begin{thebibliography}{99}

\bibitem{AguilarSaavedra:2012va} 
  J.~A.~Aguilar-Saavedra and A.~Juste,
  arXiv:1205.1898 [hep-ph].

\bibitem{Aaltonen:2011kc} 
  T.~Aaltonen {\it et al.}  [CDF Collaboration],
  Phys.\ Rev.\ D {\bf 83}, 112003 (2011).

\bibitem{CDF8.7} 
  T.~Aaltonen {\it et al.}  [CDF Collaboration],
  CDF note 10807 (2012).

\bibitem{Abazov:2011rq} 
  V.~M.~Abazov {\it et al.}  [D0 Collaboration],
  Phys.\ Rev.\ D {\bf 84}, 112005 (2011).



\bibitem{Kuhn:1998kw}
 J.~H.~K\"uhn and G.~Rodrigo,
 Phys.\ Rev.\  D {\bf 59}, 054017  (1999).

\bibitem{Antunano:2007da}
 O.~Antunano, J.~H.~K\"uhn and G.~V.~Rodrigo,
 Phys.\ Rev.\  D {\bf 77},  014003  (2008).

\bibitem{Almeida:2008ug}
 L.~G.~Almeida, G.~Sterman and W.~Vogelsang,
 Phys.\ Rev.\  D {\bf 78}, 014008  (2008).

\bibitem{Bernreuther:2010ny}
 W.~Bernreuther and Z.~-G.~Si,
 Nucl.\ Phys.\ B {\bf 837}, 90 (2010).

\bibitem{Kidonakis:2011zn}
 N.~Kidonakis,
 Phys.\ Rev.\ D {\bf 84}, 011504 (2011).

\bibitem{Ahrens:2011uf}
 V.~Ahrens, A.~Ferroglia, M.~Neubert, B.~D.~Pecjak and L.~L.~Yang,
 Phys.\ Rev.\ D {\bf 84}, 074004 (2011).

\bibitem{Hollik:2011ps}
 W.~Hollik and D.~Pagani,
 Phys.\ Rev.\ D {\bf 84}, 093003 (2011).

\bibitem{Kuhn:2011ri}
 J.~H.~K\"uhn and G.~Rodrigo,
 JHEP {\bf 1201}, 063 (2012).

\bibitem{Campbell:2012uf}
 J.~M.~Campbell and R.~K.~Ellis,
 arXiv:1204.1513 [hep-ph].

\bibitem{Bernreuther:2012sx}
 W.~Bernreuther and Z.~-G.~Si,
 Phys.\ Rev.\ D {\bf 86}, 034026 (2012).












  
\bibitem{Djouadi:2009nb} 
  A.~Djouadi, G.~Moreau, F.~Richard and R.~K.~Singh,
  Phys.\ Rev.\ D {\bf 82}, 071702 (2010);
  P.~H.~Frampton, J.~Shu and K.~Wang,
  Phys.\ Lett.\ B {\bf 683}, 294 (2010);
Q.~-H.~Cao, D.~McKeen, J.~L.~Rosner, G.~Shaughnessy and C.~E.~M.~Wagner,
  Phys.\ Rev.\ D {\bf 81}, 114004 (2010);
  Y.~Bai, J.~L.~Hewett, J.~Kaplan and T.~G.~Rizzo,
  JHEP {\bf 1103}, 003 (2011).
  M.~Cvetic, J.~Halverson and P.~Langacker,
  arXiv:1209.2741 [hep-ph].


\bibitem{Jung:2009jz} 
  S.~Jung, H.~Murayama, A.~Pierce and J.~D.~Wells,
  Phys.\ Rev.\ D {\bf 81}, 015004 (2010);
  S.~Jung, A.~Pierce and J.~D.~Wells,
  Phys.\ Rev.\ D {\bf 83}, 114039 (2011).


\bibitem{Cheung:2009ch}
  K.~Cheung, W.~-Y.~Keung and T.~-C.~Yuan,
  Phys.\ Lett.\ B {\bf 682} (2009) 287;
  K.~Cheung and T.~-C.~Yuan,
  Phys.\ Rev.\ D {\bf 83}, 074006 (2011);
  J.~Shelton and K.~M.~Zurek,
  Phys.\ Rev.\ D {\bf 83}, 091701 (2011).
  
\bibitem{Nelson:2011us} 
  A.~E.~Nelson, T.~Okui and T.~S.~Roy,
  Phys.\ Rev.\ D {\bf 84}, 094007 (2011);
  J.~A.~Aguilar-Saavedra and M.~P\'erez-Victoria,
  JHEP {\bf 1109}, 097 (2011);
  K.~Blum, Y.~Hochberg and Y.~Nir,
  JHEP {\bf 1110}, 124 (2011);
  P.~Ko, Y.~Omura and C.~Yu,
  JHEP {\bf 1201}, 147 (2012)
  
  
\bibitem{Shu:2009xf} 
  J.~Shu, T.~M.~P.~Tait and K.~Wang,
  Phys.\ Rev.\ D {\bf 81}, 034012 (2010);
  A.~Arhrib, R.~Benbrik and C.~-H.~Chen,
  Phys.\ Rev.\ D {\bf 82}, 034034 (2010);
  I.~Dorsner, S.~Fajfer, J.~F.~Kamenik and N.~Kosnik,
  Phys.\ Rev.\ D {\bf 81}, 055009 (2010);
  I.~Dorsner, S.~Fajfer, J.~F.~Kamenik and N.~Kosnik,
  Phys.\ Rev.\ D {\bf 82}, 094015 (2010);
  Z.~Ligeti, G.~M.~Tavares and M.~Schmaltz,
  JHEP {\bf 1106}, 109 (2011);
  B.~Grinstein, A.~L.~Kagan, M.~Trott and J.~Zupan,
  Phys.\ Rev.\ Lett.\  {\bf 107}, 012002 (2011).


\bibitem{Barcelo:2011vk} 
  R.~Barcelo, A.~Carmona, M.~Masip and J.~Santiago,
  Phys.\ Lett.\ B {\bf 707}, 88 (2012);
  G.~M.~Tavares and M.~Schmaltz,
  Phys.\ Rev.\ D {\bf 84}, 054008 (2011);
  E.~Alvarez, L.~Da Rold, J.~I.~S.~Vietto and A.~Szynkman,
  JHEP {\bf 1109}, 007 (2011);
  J.~A.~Aguilar-Saavedra and M.~P\'erez-Victoria,
  Phys.\ Lett.\ B {\bf 705}, 228 (2011).


\bibitem{Cao:2011ew} 
  J.~Cao, L.~Wang, L.~Wu and J.~M.~Yang,
  Phys.\ Rev.\ D {\bf 84}, 074001 (2011);
  E.~L.~Berger, Q.~-H.~Cao, C.~-R.~Chen, C.~S.~Li and H.~Zhang,
  Phys.\ Rev.\ Lett.\  {\bf 106}, 201801 (2011);
  B.~Bhattacherjee, S.~S.~Biswal and D.~Ghosh,
  Phys.\ Rev.\ D {\bf 83}, 091501 (2011);
  J.~A.~Aguilar-Saavedra and M.~Perez-Victoria,
  Phys.\ Lett.\ B {\bf 701}, 93 (2011);
  J.~A.~Aguilar-Saavedra and J.~Santiago,
  Phys.\ Rev.\ D {\bf 85}, 034021 (2012).

\bibitem{Gresham:2011pa} 
  M.~I.~Gresham, I.~-W.~Kim and K.~M.~Zurek,
  Phys.\ Rev.\ D {\bf 83}, 114027 (2011);
  J.~L.~Hewett, J.~Shelton, M.~Spannowsky, T.~M.~P.~Tait and M.~Takeuchi,
  Phys.\ Rev.\ D {\bf 84}, 054005 (2011);
  G.~F.~Giudice, B.~Gripaios and R.~Sundrum,
  JHEP {\bf 1108}, 055 (2011);
  U.~Haisch and S.~Westhoff,
  JHEP {\bf 1108}, 088 (2011);
  S.~Jung, A.~Pierce and J.~D.~Wells,
  Phys.\ Rev.\ D {\bf 84}, 091502 (2011);
  E.~L.~Berger, Q.~-H.~Cao, J.~-H.~Yu and C.~-P.~Yuan,
  Phys.\ Rev.\ D {\bf 84}, 095026 (2011).

\bibitem{Diener:2009ee} 
  R.~Diener, S.~Godfrey and T.~A.~W.~Martin,
  Phys.\ Rev.\ D {\bf 80}, 075014 (2009).

\bibitem{Aad:2012ug} 
  G.~Aad {\it et al.}  [ATLAS Collaboration],
  Eur.\ Phys.\ J.\ C {\bf 72}, 2039 (2012).

\bibitem{Chatrchyan:2011hk} 
  S.~Chatrchyan {\it et al.}  [CMS Collaboration],
  Phys.\ Lett.\ B {\bf 709}, 28 (2012).
  
\bibitem{Chatrchyan:2012xv} 
  S.~Chatrchyan {\it et al.}  [CMS Collaboration],
  Phys.\ Lett.\ B {\bf 717}, 129 (2012).

\bibitem{Drobnak:2012cz} 
  J.~Drobnak, J.~F.~Kamenik and J.~Zupan,
  Phys.\ Rev.\ D {\bf 86}, 054022 (2012).

\bibitem{Alvarez:2012ca} 
  E.~Alvarez and E.~C.~Leskow,
  arXiv:1209.4354 [hep-ph].

\bibitem{Drobnak:2012rb} 
  J.~Drobnak, A.~L.~Kagan, J.~F.~Kamenik, G.~Perez and J.~Zupan,
  arXiv:1209.4872 [hep-ph].

\bibitem{AguilarSaavedra:2011hz} 
  J.~A.~Aguilar-Saavedra and M.~Perez-Victoria,
  Phys.\ Rev.\ D {\bf 84}, 115013 (2011).
  
\bibitem{Frixione:2008ym}
 S.~Frixione and B.~R.~Webber,
 arXiv:0812.0770 [hep-ph].

\bibitem{Nadolsky:2008zw}
 P.~M.~Nadolsky, H.~-L.~Lai, Q.~-H.~Cao, J.~Huston, J.~Pumplin,
D.~Stump, W.~-K.~Tung and C.~-P.~Yuan,
 Phys.\ Rev.\ D {\bf 78},  013004 (2008).
 
 
\end{thebibliography}
\end{document}